# *MorphoSim*: An efficient and scalable phase-field framework for accurately simulating multicellular morphologies

Xiangyu Kuang[1†], Guoye Guan[1†], Chao Tang[1,2,3*], Lei Zhang[1,4*]

[1] *Center for Quantitative Biology, Peking University, Beijing 100871, China*

[2] *Peking-Tsinghua Center for Life Sciences, Peking University, Beijing 100871, China*

[3] *School of Physics, Peking University, Beijing 100871, China*

[4] *Beijing International Center for Mathematical Research, Peking University, Beijing 100871, China*

[†] *These authors contributed equally to this work.*

[*] *For correspondence: tangc@pku.edu.cn (CT), zhangl@math.pku.edu.cn (LZ)*

**Abstract**

The phase field model can accurately simulate the evolution of microstructures with complex morphologies, and it has been widely used for cell modeling in the last two decades. However, compared to other cellular models such as the coarse-grained model and the vertex model, its high computational cost caused by three-dimensional spatial discretization hampered its application and scalability, especially for multicellular organisms. Recently, we built a phase field model coupled with *in vivo* imaging data to accurately reconstruct the embryonic morphogenesis of *Caenorhabditis elegans* from 1- to 8-cell stages [Kuang et al, *PLoS Comput. Biol.*, 2022]. In this work, we propose an improved phase field model by using the stabilized numerical scheme and modified volume constriction. Then we present a scalable phase-field framework, *MorphoSim*, which is 100 times more efficient than the previous one, and can simulate over 100 mechanically interacting cells. Finally, we demonstrate how *MorphoSim* can be successfully applied to reproduce the assembly, self-repairing, and dissociation of a synthetic artificial multicellular system - the *synNotch* system.

## 1. Introduction

The multicellular systems formed by physically interacting cells are widespread in animals, plants, and microorganisms like fungus and choanoflagellate [2-5]. They usually consist of multiple cell types and characteristic spatial organizations, ranging in scales from tissue, organ, to an individual, and are often involved with embryogenesis and organogenesis [6-7]. For example, the mouse valley-like small intestine, which contains ~ 250 cells with 6 cell types in specific proportions, plays a role in nutrient uptake and has strong architectural robustness and regenerative capacity [8-10]. The mechanics of interacting cell aggregates, as well as their morphological and morphogenetic effects, has attracted increasing attention in the fields of cell biology, developmental biology, cancer biology, etc [11-13]. Apart from the natural ones, many engineered systems have been constructed *in vitro* artificially, such as the organoid and embryoid used for developmental biology research and high-throughput drug test [14-16]. Furthermore, synthetic biologists are trying to program the cellular interaction *de novo* to build customized multicellular living machines, robots, or patterns, using bottom-up or top-down engineering approaches. *synNotch* and *Xenobot* systems are two cutting-edge representatives, which are constructed for designed functions using hundreds of animal cells and have broad application areas like medical treatment and synthetic development [17-20]. An urgent need for a computational tool that can efficiently and accurately simulate multicellular morphological behaviors is emerging [21,22].

Many physical models have been utilized to study the morphological dynamics of multicellular systems, including cellular automaton [23], cellular Potts model [24], coarse-grained model [25], Voronoi tessellation model [26], vertex model [27], multi-particle model [28], phase field model [29], etc. These models have provided much mechanistic knowledge and insights for understanding the biological processes. However, simulation results of the same system may vary from model to model – when a precise description of the real system is critical the accuracy of the model output becomes an important issue [30]. Besides, most of the existing models lack a comprehensive description of cell shape and interface-based cell-cell interaction, and often lack detailed validation by quantitative comparison with experiments, especially at three-dimensional (3D) and single-cell level. Recently, a phase field model was established with the help of 3D time-lapse imaging data of *C. elegans* embryogenesis from 1- to 8-cell stages, in which cell surface tension, cell-eggshell and cell-cell repulsion, cell-cell attraction, and cell volume constriction were considered (Fig. 1(a)) [1]. The model can accurately capture the cell morphology *in vivo* and infer the underlying biophysical properties such as intercellular adhesion (Fig. 1(b)). Despite its outstanding precision, the model was limited by the high computational cost caused by 3D spatial discretization and increasing cell numbers, especially when the cell number reaches dozens and hundreds.

In this work, we propose an efficient and scalable phase-field framework that can accurately simulate multicellular morphologies. We first develop a stabilized numerical scheme that allows for large time steps. Next, we enhance the precision of cell volume control to avoid "cell disappearance". Finally, a phase-field framework, *MorphoSim*, is established along with Matlab-based software. By testing on the simulations of *C. elegans* embryogenesis, it can achieve a computational efficiency more than 100 times of the previous one and is capable of computing over 100 cells. As an application, the *MorphoSim* framework can successfully reproduce the assembly, self-repairing, and dissociation of the *synNotch* system reported in [17].

## 2. Phase field model and computation

*2.1. Review of the phase field model*

The original phase field model adopted by this study considered the surface tension $F_{\text{ten}}$ and volume constriction $F_{\text{vol}}$ imposed on a cell, and the repulsion $F_{\text{rep}}$ and attraction $F_{\text{atr}}$ between cells. The spatial constraint from the eggshell is also repulsive to a cell toward the embryo's center and thus included in the term $F_{\text{rep}}$ [1]. A cell $i$ with prescribed volume $V_i(t)$ is represented by a phase field $\phi_i(\mathbf{r}, t)$ and assumed to deform and migrate in an overdamped domain $\Omega$, following the governing equations:

$$F_{\text{ten}} = -\gamma \left( \Delta \phi_i - cW'(\phi_i) \right) \frac{\nabla \phi_i}{|\nabla \phi_i|^2}, \quad (1)$$

$$F_{\text{vol}} = M \left( \int_\Omega \phi_i \mathrm{d}\mathbf{r} - V_i(t) \right) \hat{\mathbf{n}}, \quad (2)$$

$$F_{\text{rep}} = \left( g_e \phi_i \phi_e^2 + g \phi_i \sum_{j \neq i}^{N} \phi_j^2 \right) \frac{\nabla \phi_i}{|\nabla \phi_i|^2}, \quad (3)$$

$$F_{\text{atr}} = \sum_{j \neq i}^{N} \sigma_{i,j} \nabla \phi_j, \quad (4)$$

$$\frac{\partial \phi_i}{\partial t} = -\frac{1}{\tau} \left( F_{\text{ten}} + F_{\text{vol}} + F_{\text{rep}} + F_{\text{atr}} \right) \cdot \nabla \phi_i. \quad (5)$$

Here $\Delta$ and $\nabla$ are Laplacian and gradient operators respectively; $\gamma$ denotes the cell surface tension that drives the cell shape to be spherical; $c$ controls the cell boundary thickness; $W(\phi) = \phi^2(\phi - 1)^2$ separates two phases at $\phi = 0$ and $\phi = 1$, corresponding to the exterior and interior of a cell; $g_e$ and $g$ represent the eggshell-cell and cell-cell repulsions respectively; $\sigma_{i,j}$ represents the attraction strength between the *i*-th and *j*-th cells; $M$ is the volume constriction strength; $\hat{\mathbf{n}}$ is the unit normal vector at the interface and orients inward; $\tau$ is the ambient viscosity.

During *C. elegans* embryogenesis, every cell has unique and identifiable developmental behavior and is systematically named based on its cell type, lineal origin, and spatial location [31]. With the input of cell division order and axis and volume segregation ratio measured experimentally, the previous phase-field framework successfully reproduced the typical embryonic morphology up to the 8-cell stage (Fig. 1(a) and (b)). The conserved cell-cell contact map was fully reproduced and the cell-cell adhesion programs were reversely inferred, including the low adhesion in EMS-P2 contact and ABpl's posterior reported before [32,33]. To distinguish the time scales with different meanings, hereafter the time in experiment and computer are referred to as "*in vivo* time" and "*in silico* time" (*in silico* time = step number × step length = $n_t \delta t$) respectively, while the time cost for simulation is termed "computing time" and is one of the optimization targets in this work.

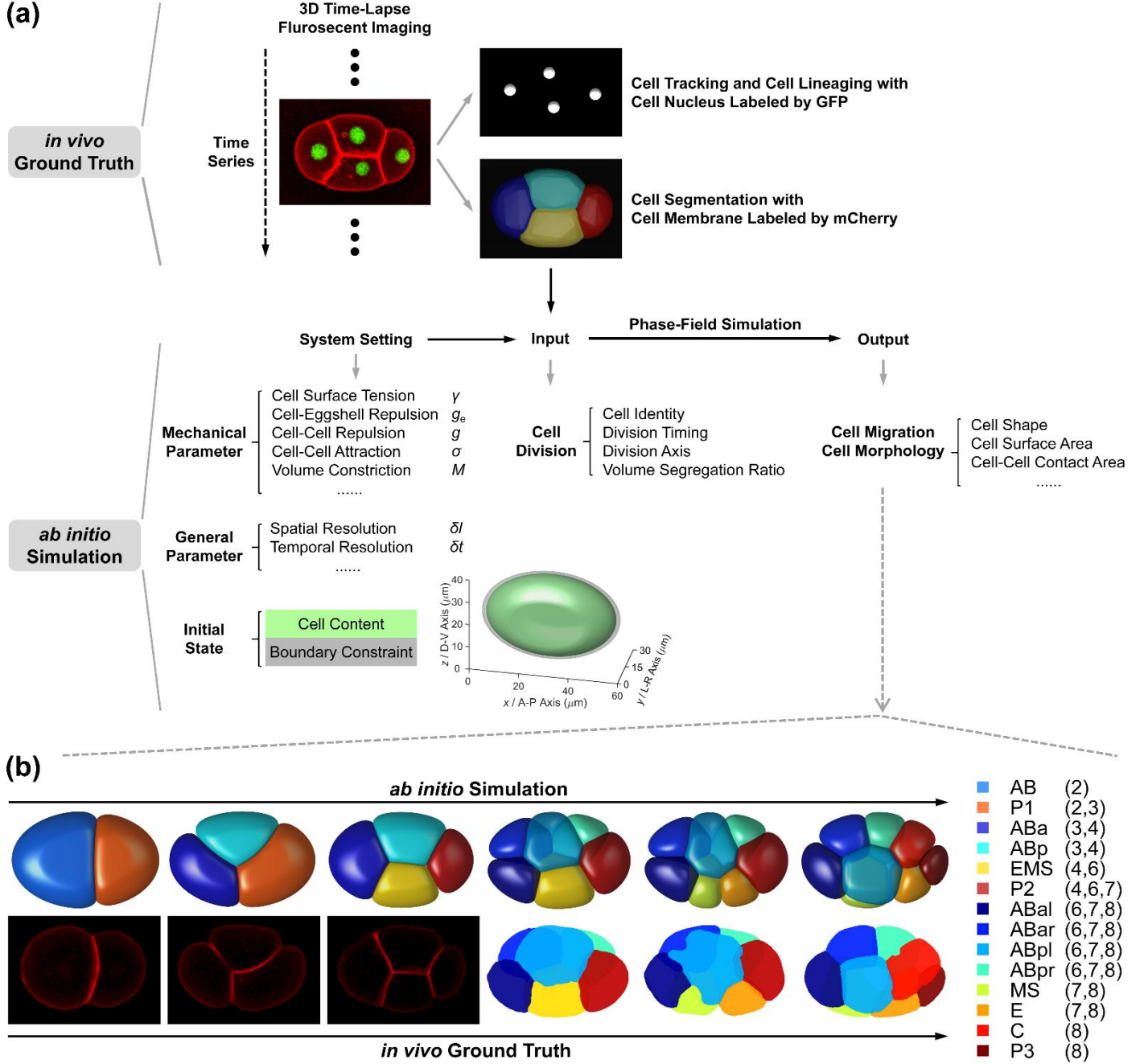

**Fig. 1.** The phase-field framework for *C. elegans* early embryogenesis. (a) Flow chart. (b) Comparison between *ab initio* simulation and *in vivo* ground truth of *C. elegans* embryonic morphologies from 2- to 8-cell stages. The embryonic structures at 2-, 3-, and 4-cell stages are 2-dimensional and shown by fluorescent images; the ones at 6-, 7-, and 8-cell stages are 3-dimensional and shown by segmented images. The cells' identities and corresponding colors and existing stages (total cell number) are denoted on right. All the subfigures, except the schematics for the initial state setting in (a), are adapted from [1] with granted permission.

## 2.2. Stabilized numerical scheme

The high computational cost of phase-field simulation is largely attributed to its 3D spatial discretization and cell number increase. One direct attempt to reduce the computational cost is to minimize the spatial and temporal resolutions while guaranteeing the results are consistent with the previous biological findings [1]. The requirements that judge if the framework works precisely enough from the 1- to 8-cell stages mainly include 3 parts: 1. the cell-cell contact maps are the same as the ones conserved between individual embryos (Fig. S1(a)-(d)); 2. the previously reported cell-cell adhesion programs can be inferred by comparison to experiment and parameter scanning; 3. the embryonic morphologies resemble the ones *in vivo*. The criteria are quantified and detailly introduced in Supplementary Note 1. Given the original spatial discretization $\delta l = 0.25$ (μm) and temporal discretization $\delta t = 0.10$, we scan the spatial and temporal discretization and repeat the simulation procedure from 1- to 8-cell stages as shown in Fig. S2(a). We find that three requirements can be satisfied by a coarser spatial discretization $\delta l = 0.50$ (μm) and time step $\delta t = 0.30$ (Supplementary Note 1). However, such improvement is very limited.

The stability restriction on the time step impedes the further acceleration of computation. In order to allow a much larger time step than the explicit schemes, we first adopt a first-order semi-implicit scheme [34]:

$$\frac{\tau}{\delta t}(\phi_i^{n+1} - \phi_i^n) = \gamma \Delta \phi_i^{n+1} + F_i^n, \tag{6}$$

where $n$ denotes the $n$-th time step. The linear term $\gamma \Delta \phi_i^{n+1}$ was treated implicitly while the nonlinear term $F_i^n$ is treated explicitly and expressed by:

$$F_i^n = -\gamma c\big(4(\phi_i^n)^3 - 6(\phi_i^n)^2 + 2\phi_i^n\big) - g_e \phi_i^n \phi_e^2 - g\phi_i^n \sum_{j \neq i}^N (\phi_j^n)^2 - \nabla \phi_i^n \cdot \sum_{j \neq i}^N \sigma_{i,j} \nabla \phi_j^n + M\left(V_i(t) - \int_\Omega \phi_i^n d\boldsymbol{r}\right)|\nabla \phi_i^n|, \tag{7}$$

Next, we relax the restriction by the stabilization method. The main idea is to add an artificial stabilization term that has a dissipative effect to balance the instability caused by the explicit treatment of the nonlinear term [35,36]. Here, we first introduce a first-order stabilization term $-S(\phi_i^{n+1} - \phi_i^n)$ to alleviate the strict constraint on the temporal evolution:

$$\frac{\tau}{\delta t}(\phi_i^{n+1} - \phi_i^n) = \gamma \Delta \phi_i^{n+1} + F_i^n - S(\phi_i^{n+1} - \phi_i^n), \tag{8}$$

where $S$ is a positive coefficient proportional to the dissipative effect. Increasing $S$ will make computation stabler but also introduce extra numerical error. Thus, it is necessary to keep a balance between stability and accuracy. Therefore, we optimize $S$ by finding its minimal value that can stabilize a simulation with $\delta t > 0.3$ while preserving enough accuracy, i.e., still reconstructing the morphogenetic dynamics observed experimentally (Fig. 1(b)). For each $\delta t$ from 0.3 to 2.0 in a step of 0.1, we perform simulation from 1- to 8-cell stages to search for the optimal $S$ value in a step of 0.1. Here, the time scale of each simulation is proportionally fitted to the one without stabilization term ($\delta t = 0.3$; baseline), using the duration of 6- and 7-cell stages determined by their first quasi-steady states; then the *in silico* time for 8-cell stage is calculated with the linear relationship given the base value set as 15000. It's shown that the system bifurcates into another 8-cell topology when $\delta t$ exceeds 1.8 (Fig. S3(a)). Besides, with the increment of $\delta t$, the computing time is not always monotonously decreasing (Fig. 2(a)), revealing that the *in silico* time $n_t \delta t$ for the same stage/process also increases (Fig. S3(b) and Table S1). This overdamping-like effect, i.e., the increase of *in silico* time $n_t \delta t$, was also shown in previous research and may be caused by the numerical error of the stabilization term in the first-order scheme, which then limits the computing time reduction gained from a larger $\delta t$ [37].

To solve the problem above, we further introduce a semi-implicit scheme with a second-order stabilization term to achieve less numerical error:

$$\frac{\tau}{2\delta t}\big(3\phi_i^{n+1} - 4\phi_i^n + \phi_i^{n-1}\big) = \gamma \Delta \phi_i^{n+1} + 2F_i^n - F_i^{n-1} - S\big(\phi_i^{n+1} - 2\phi_i^n + \phi_i^{n-1}\big). \tag{9}$$

We extend the scanning range of $\delta t$ to 0.3 ~ 2.5 and adopt the same optimization procedure for finding the optimal $S$ value. The second-order scheme takes full advantage of the larger $\delta t$ by avoiding the overdamping-like effect, therefore, substantially reducing the computing time and keeping the simulations with different $\delta t$ values scalable (Fig. 2(b)). The perfect scalability is suddenly broken when $\delta t$ reaches 2.1, probably due to numerical error accumulation. Thus, we choose the temporal resolution $\delta t = 2.0$ with $S = 12$ as the optimal condition for further simulation, which can recapitulate the 1- to 8-cell morphogenesis of *C. elegans* embryo (Fig. 3(c)). Although the second-order scheme gains an edge over the first-order one in computing time, it's worth pointing out that the first-order scheme is better at maintaining numerical stability [36]. In other words, the first-order scheme allows larger values assigned to the parameters. Note that the $\delta t$ and $S$ values presented here are optimized to minimize the computing time, and one may select other values according to the actual problems or parameter settings.

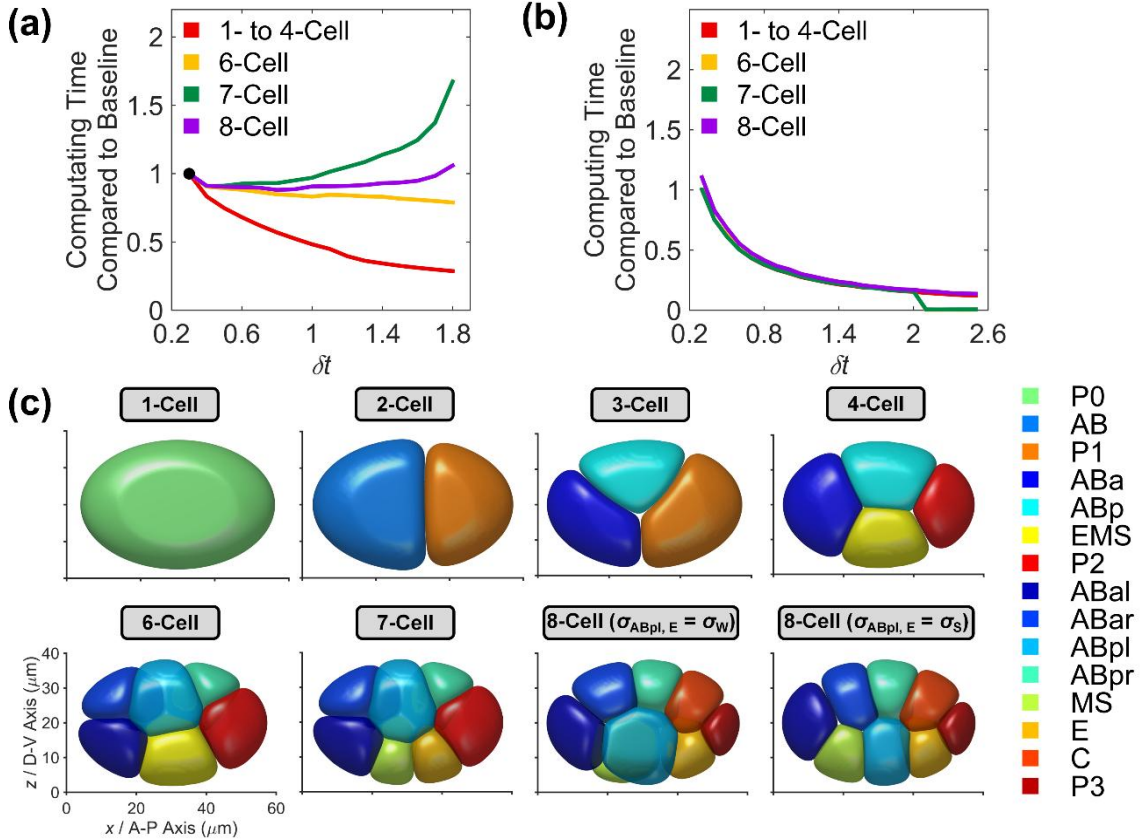

**Fig. 2.** Framework improvement by the addition of stabilization term. (a) Computing time compared to the baseline at different stages in the first-order scheme. (b) Computing time compared to the baseline at different stages in the second-order scheme. In both (a) and (b), the baseline is set as the computing time when $\delta t = 0.3$ and no stabilization term is added. (c) Embryonic morphologies from 1- to 8-cell stages in the second-order scheme with $\delta t = 2.0$. The cells' identities and corresponding colors are denoted on right; $\sigma_S$ and $\sigma_W$ denote the relatively strong and weak cell-cell attraction inferred at 4-cell stage respectively (Supplementary Note 1).

*2.3. A new formation of volume constriction to avoid "cell disappearance"*

After the improvement of the numerical scheme, we then perform simulations for the later stages of *C. elegans* embryogenesis to find the maximum cell number afforded by the current framework. With the experimentally-measured cell division order and axis and volume segregation ratio inputted, the simulation proceeds properly until the 24-cell stage, when the whole phase field of P4 cell, $\phi_{P4}$, shrinks to zero erroneously (Fig. 3(a) and Movie S1). Hereafter, such a phenomenon caused by numerical error is referred to as "cell disappearance" in this work. In the simulations, cell disappearance is quantitatively defined when a cell's phase field is globally smaller than 0.5.

When cell disappearance occurs, the volume constriction (Eq. (2)) fails to prevent a cell's phase field from dropping to the homogeneous state $\phi \equiv 0$. Further simulation on a single cell with different sizes and without any external force reveals that a threshold of cell size exists to determine if cell disappearance happens (Fig. 3(b)). During the relaxation of a free cell, its interior always shrinks, while the width of diffusing interface remains nearly constant. If the cell size is below the threshold ($R \leq 4$ μm or $V \leq 268$ μm³), the interior of phase field shrinks to disappear and its boundaries gradually overlap, making the phase field eventually converge to $\phi \equiv 0$, namely, cell disappearance (Fig. 3(b)). As a small cell size appears in all kinds of biological processes *in vivo*, like blastomere cleavage and cell apoptosis, solving the problem of cell disappearance is vital for simulations of those scenes [38,39].

The phenomenon of cell disappearance can be avoided in two ways. First, we replace the volume constriction (Eq. (2)) with a new formation based on the relative error of volume, instead of the absolute error controlled in the previous one:

$$F_{vol} = M' \left( \frac{\int_\Omega \phi_i d\mathbf{r}}{V_i(t)} - 1 \right) \hat{\mathbf{n}}. \quad (10)$$

where $M'$ is the volume constriction strength. By limiting the relative error, the new formation can achieve a more accurate simulated volume, consistent with the prescribed one measured experimentally or designated arbitrarily. Apart, the stronger volume constriction for the cells with small size ($\sim 10^2$ μm³) lowers the cell size threshold of cell disappearance. The new formation also maintains numerical stability when handling the relatively larger cells ($10^3 \sim 10^4$ μm³), while strengthening volume constriction in the previous formation (i.e., by amplifying $M$) would cause numerical instability easily. The second way to resist cell disappearance is to amplify the value of parameter $c$, the positive coefficient of double-well potential $W(\phi)$, so that the ability of a phase field to separate into two phases can be enhanced (Eq. (1)) (Fig. 3(c)). Here, we select $c = 2$, $\delta t = 1.5$, and $M' = 8$ (with a similar error level to the previous formation at 4-cell stage) considering both numerical stability and accuracy for simulations of later stages. The final iterative process with the stabilization term added and the volume constriction modified is detailed in Supplementary Note 2.

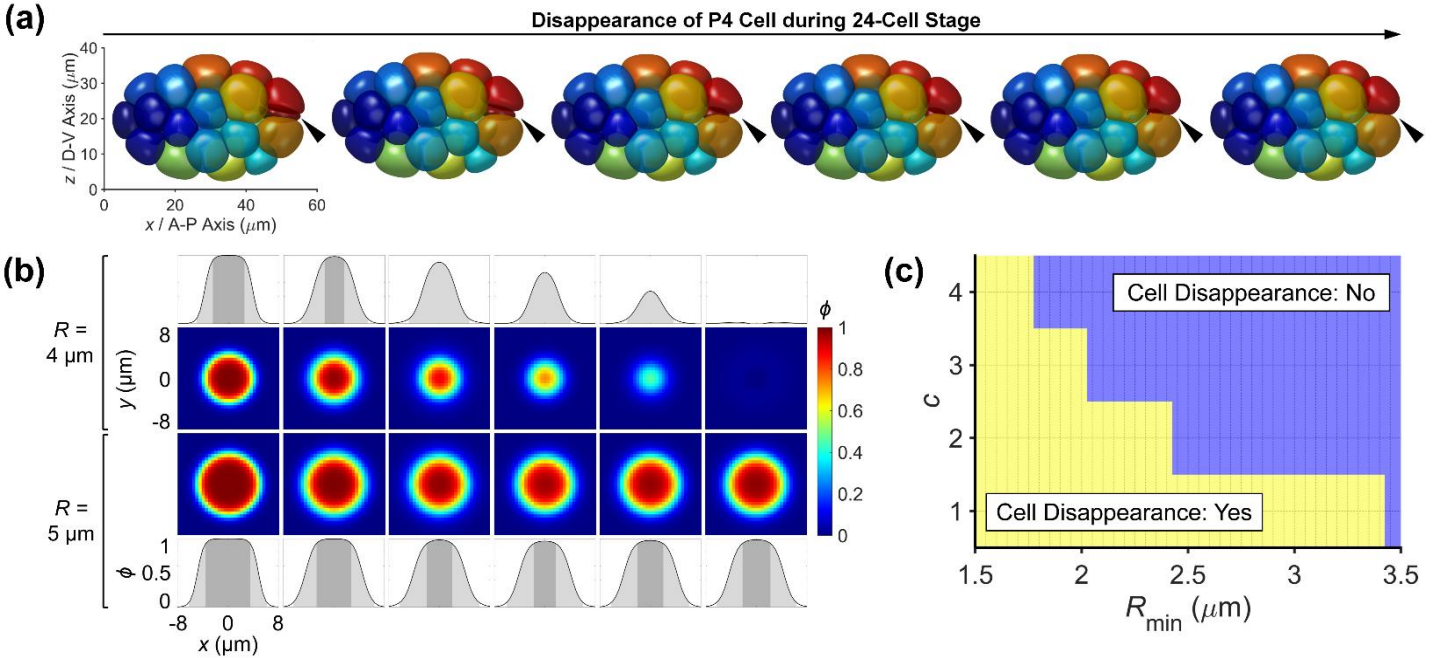

Fig. 3. Cell disappearance that happens when the cell size is too small. (a) Gradual disappearance of the P4 cell during 24-cell stage; the P4 cell is colored dark red and pointed by an arrowhead. (b) Phase field distribution in 1D (*x* axis) and 2D (*xy* plane) of a free cell with a radius of 4 μm (upper two rows) and 5 μm (lower two rows). The single-cell simulation is conducted with open boundaries and illustrated with an *in silico* interval of 20 from left to right. In the 1D distribution, the boundary of a cell ($0.07 \leq \phi < 0.93$) is painted with light gray while the interior ($\phi \geq 0.93$) and exterior ($\phi < 0.07$) of a cell are painted with dark gray and white respectively. (c) A screening on parameter *c* with different cell radii $R_{min}$; yellow and blue indicate if the cell disappears or not in single-cell simulation.

## 2.4. A Matlab-based GUI for automatic computation and structural illustration

Given the phase field model and numerical methods, we achieve the final framework named *MorphoSim* (<u>Morpho</u>logy <u>Sim</u>ulator). We further pack it into an open-source graphical user interface (GUI) using the software Matlab (Fig. 4(a)) [40]. One can input the binary distribution of cells, assign a cell-cell attraction matrix arbitrarily, and set up the *in silico* time, step length (i.e., temporal resolution $\delta t$), and saving interval for simulation. It should be pointed out that the eggshell boundary can be removed as an open-boundary condition and the computation can be performed on a CPU or GPU according to user requirements. When the simulation is over, the user can import the output file of a specific time point and plot the 3D structure automatically. A detailed guidebook can be seen in Supplementary Note 3.

During the 8-cell *C. elegans* embryogenesis, the weaker adhesion in ABpl-E contact was reported to be critical for the robust formation of 3D embryonic structure, which serves as a criterion to check if the framework preserves its precision (Fig. 2(c)) [1]. The input for 8-cell simulations with strong and weak adhesion in ABpl-E contact are shown in Fig. 4(b) and (c), and the initial state of phase fields is shown in Fig. 4(d). The simulation results are in line with previous findings (Fig. 4(e) and (f)) [1,33]. Using a personal computer, this simulation lasts for less than 1.5 hours on CPU (Intel(R) Core(TM) i5-10210U CPU @ 1.60GHz 2.11 GHz) and less than 12.5 minutes on GPU (NVIDIA GeForce GTX 1060) for an *in silico* time = 15000.

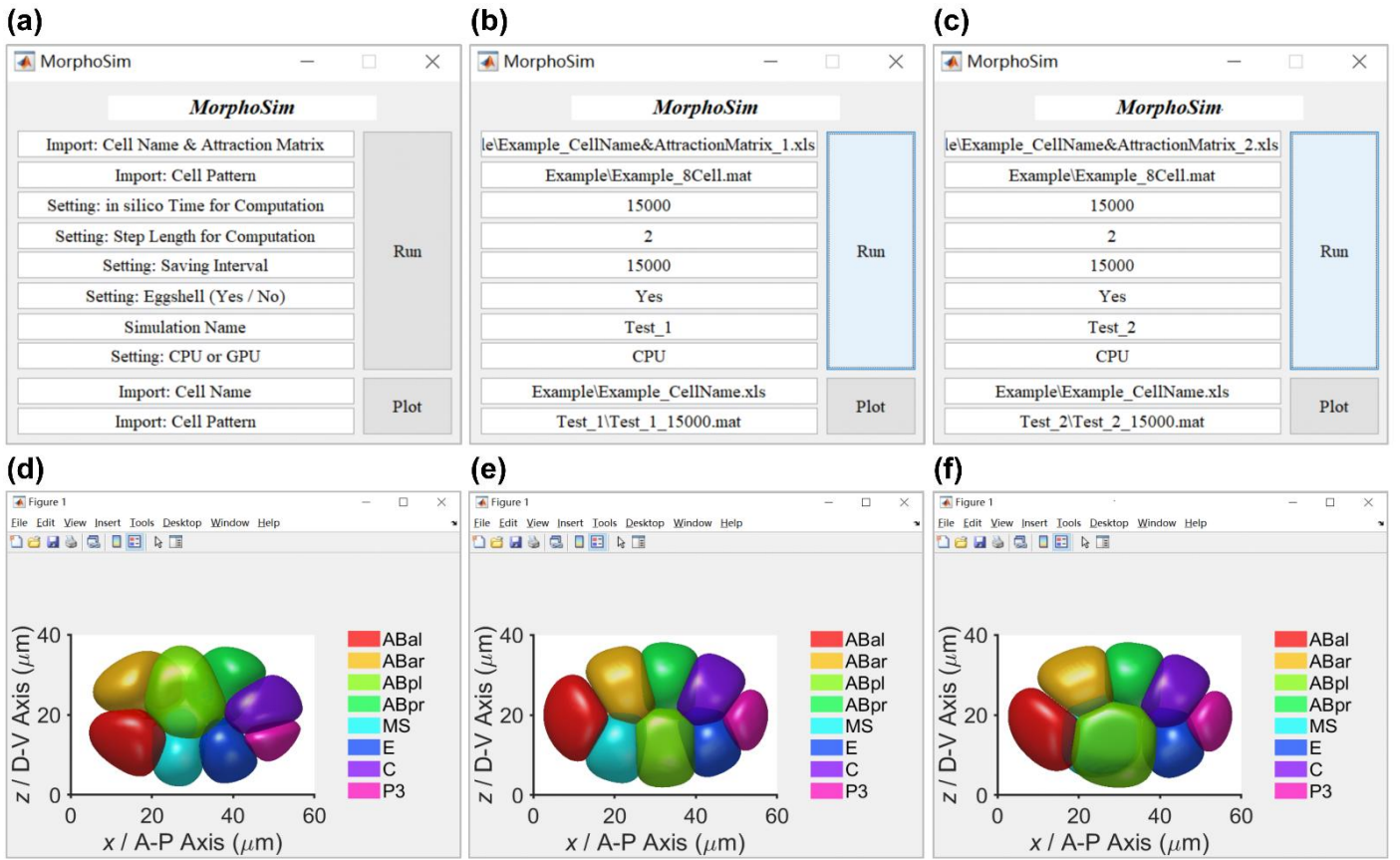

**Fig. 4.** The graphical user interface of *MorphoSim*. (a) The interface and instruction of inputs required. (b)(c) The simulation inputs for 8-cell *C. elegans* embryogenesis with strong and weak adhesion in ABpl-E contact respectively. (d) The initial state (*in silico* time = 0) of the 8-cell embryo. (e)(f) The final state (*in silico* time = 15000) of the 8-cell embryo with strong and weak adhesion in ABpl-E contact respectively.

## 3. Test and application of the *MorphoSim*

*3.1. Simulation of the C. elegans embryogenesis from 1- to 102-cell stages*

To test the availability of *MorphoSim*, we first perform the simulations of *C. elegans* embryo from 1- to over 102-cell stages according to the cell division order observed *in vivo*. For simplicity, here we consider the 6 founder cells (i.e., AB, MS, E, C, D, and P4) as well as their ancestors (i.e., P0, P1, P2, P3, and EMS) and progenies. The cell divisions in the same generation of a founder cell are pseudo-synchronous with a slight variation in reality and are idealized as a whole division group with the same cell cycle length in simulation. We use the experimentally-measured shortest cell cycle length within the group as the common value [41]; finally, we obtain the cell division order formed by 24 independent division groups as shown in Fig. 5(a). Besides, the inputted cell division axis and volume segregation ratio are acquired from [42]. The simulation from 1- to 8-cell stages is performed following the pipeline in Supplementary Note 1 while the one after 8-cell stage is set to reach a mechanical equilibrium for each stage and then the next cell division(s) would be activated; after 8-cell stage, the cell-cell attraction matrix is still simplified as binary and follows the rule that attraction between sisters cells and between non-sister cells are relatively weak and strong respectively [1]. With the inputs above, the phase-field framework successfully simulates a multicellular system with up to 102 cells, without any cell disappearance (Fig. 5(b) and Movie S2).

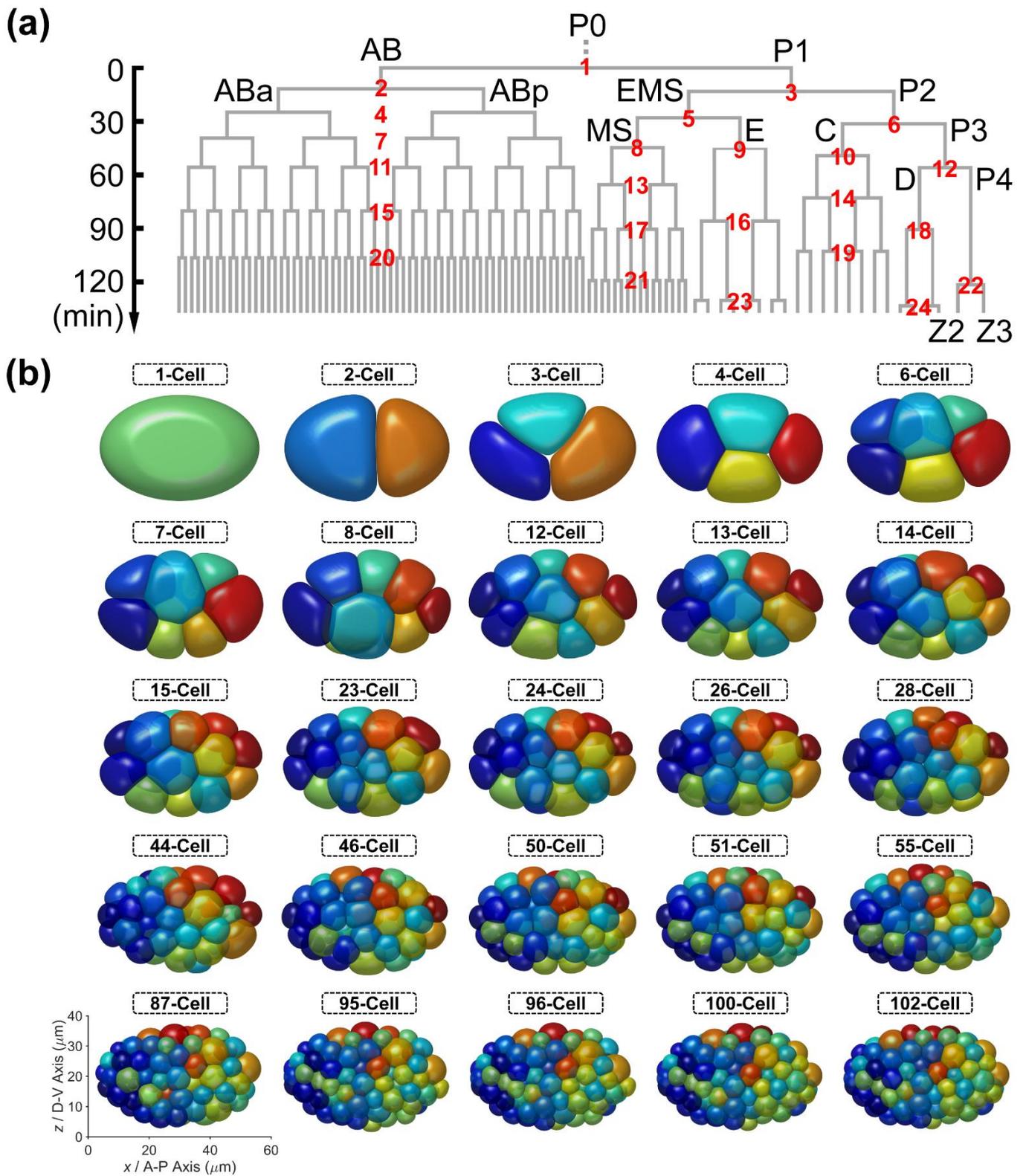

**Fig. 5.** Simulation from 1- to 102-cell stages using the cell division order and axis and volume segregation ratio measured in *C. elegans* embryogenesis. (A) A cell lineage tree in which the cell divisions in the same generation and from the same founder cell (i.e., AB, MS, E, C, D, or P4) are regarded as synchronous; the cell cycle length of each division group is approximated with the shortest one among the cells obtained from experimental measurement; the cell division order is labeled near each cell division group. (B) Simulated embryonic morphologies from 1- to 102 cell stages.

Next, we test its computational efficiency in GPU and CPU compared to the original framework, whose *in silico* time scale has been well fitted with the *in vivo* one [1]. we perform the simulation from 6- to 8-cell stages. The time scales between the two frameworks are proportionally fitted using the duration of 6- and 7-cell stages determined by their first quasi-steady states so that the simulation process of each stage is equivalent in both frameworks. It shows that *MorphoSim* reaches a computational efficiency of more than $10^2$ times of the previous one averagely, where the computing time in GPU and corresponding *in vivo* time are in the same order of magnitude (Table 1). Further, we increase the cell number to 25, 50, and 100 and estimate the time cost for the system to evolve in 5 min in reality. For the original framework, the computation is out of memory in both GPU and CPU when the cell number reaches 100. The problem is well solved in *MorphoSim*, allowing full utilization of computational resources and numerical experiments coupled with parameter scanning (Table 1).

Table 1. Computing time in GPU and CPU using the previous and current frameworks.

| Cell Number | *in vivo* Time | Computing Time | | | |
|---|---|---|---|---|---|
| | | Previous Framework [1] | | Current Framework (*MorphoSim*) | |
| | | GPU | CPU | GPU | CPU |
| 6 | 1.1 min | 35.3 min | 1428.0 min | 0.3 min | 4.1 min |
| 7 | 0.8 min | 30.9 min | 1116.1 min | 0.1 min | 2.0 min |
| 8 | 5.3 min | 229.8 min | 7767.8 min | 2.6 min | 39.5 min |
| 25 | 5.0 min | 854.3 min | 26067.4 min | 17.2 min | 186.6 min |
| 50 | 5.0 min | Out of Memory | 67629.4 min | 29.8 min | 356.3 min |
| 100 | 5.0 min | Out of Memory | Out of Memory | 93.8 min | 706.1 min |

Note: The *in vivo* times of 6-, 7-, and 8-cell stages are obtained from experiments and have been proportionally fitted to the *in silico* times used for simulations. GPU, NVIDIA Tesla P100; CPU, Inter Xeno E5-2680 v3.

*3.2. Reconstruction of the assembly, self-repairing, and dissociation of the synNotch systems*

The *synNotch* system is the state-of-the-art methodology to generate self-organizing multicellular living machines, which consist of multiple cell types with genetically programmed differential adhesion [17]. The topology of cell aggregate is dependent on the combinatorial adhesive strengths and can self-repair after cleavage and dissociate after eliminating the adhesive protein. Here, we employ *MorphoSim* to reproduce the stereotypic tricomponent and biocomponent topologies reported before, which can form spherically asymmetric and symmetric patterns corresponding to two different sets of cell adhesion programs.

In the simulation of *synNotch* systems, we add the Gaussian white noise $\boldsymbol{\xi}_i(t)$ onto each cell's motion to model the stochasticity in reality. Hence, the evolution equation turns into:

$$\frac{\partial \phi_i}{\partial t} = -\frac{1}{\tau}\left(\boldsymbol{F}_{\text{ten}} + \boldsymbol{F}_{\text{vol}} + \boldsymbol{F}_{\text{rep}} + \boldsymbol{F}_{\text{atr}} + \kappa \boldsymbol{\xi}_i(t)\right) \cdot \nabla \phi_i, \tag{11}$$

where $\kappa$ is the noise strength. For all the simulations, the computational domain is set as a 128×128×128 cubic grid with spatial discretization $\delta l = 0.5$ (μm) and time step $\delta t = 1.25$, and $\kappa$ is assigned $\frac{\sqrt{5}}{10}$. It's worth noting that the spatial scale of simulation can be adjusted to the experimental one by rescaling parameters. Here we provide a rescaled parameter setting with the spatial scale of simulations approximating the experimental scales of *synNotch* system (Table S3) [17]. To set up the initial state, we uniformly randomize the cell positions inside the cubic domain, while each cell is assigned as a sphere with a radius $R = 5$ μm; then we impose an intercellular repulsion $\boldsymbol{F}_{i,j}$ and a repulsion $\boldsymbol{F}_{\text{boundary},i}$ between cell and boundary to eliminate the possible overlap between cellular regions:

$$F_{i,j} = \begin{cases} k_1(\bar{r}_i - \bar{r}_j), & |\bar{r}_i - \bar{r}_j| < 2R \\ 0, & |\bar{r}_i - \bar{r}_j| \geq 2R \end{cases}, \quad (12)$$

$$F_{\text{boundary},i} = \begin{cases} k_2(d_{\text{boundary},i} - R), & |d_{\text{boundary},i}| < R \\ 0, & |d_{\text{boundary},i}| \geq R \end{cases}. \quad (13)$$

Here, $\bar{r}_i$ and $\bar{r}_j$ are the centroids of cell $i$ and cell $j$; $d_{\text{boundary},i} = (d_{i,x}, d_{i,y}, d_{i,z}) = \min(L - \bar{r}_i, \bar{r}_i)$ is the nearest distance between the centroid of cell $i$ and the cubic boundary in three orthogonal directions; $L$ is the side length of the cubic domain; $k_1 = 0.05$; $k_2 = 0.1$. The iteration $\bar{r}_i^{n+1} = \bar{r}_i^n + F_{\text{boundary},i} + \sum_{i \neq j} F_{i,j}$ continues until all the cells have a distance no less than $2R$ to the others (roughly 200 ~ 300 time steps), then their distribution will be used as the initial condition for *synNotch* simulation. The simulation scenarios along with their parameter settings and time scales are described in Table 2.

Table 2. Parameter settings and time scales of the *synNotch* simulations.

| Topological Dynamic | Stage | Cell Type 1 | Cell Type 2 | Cell Type 3 | *in silico* Time | Computing Time (h) | Value Assignment on Cell-Cell Adhesion |
|---|---|---|---|---|---|---|---|
| Spherically Asymmetric Separation (3 Cell Types) | Start | 80 (R) | 40 (G) | 0 (B) | 50000 | 12.2 | $\sigma_{1,1} = 0.9$ $\sigma_{1,2} = 0.5$ $\sigma_{2,2} = 0.9$ |
| | End | 40 (R) | 40 (G) | 40 (B) | | | |
| Spherically Asymmetric Separation (2 Cell Types) | From Start To End | 62 (R) | 62 (G) | / | 50000 | 11.9 | |
| Spherically Symmetric Separation | | 50 (G) | 75 (B) | / | 50000 | 13.3 | $\sigma_{1,1} = 0.9$ $\sigma_{1,2} = 0.5$ $\sigma_{2,2} = 0.3$ |
| Self-Repairing | | 24 (G) | 32 (B) | / | 50000 | 4.4 | |
| Dissociation | | 50 (G) | 75 (B) | / | 50000 | 12.1 | |

Note: The painting color of each cell type in Fig. 6 is labeled in the columns "Cell Type 1", "Cell Type 2", and "Cell Type 3", where "R", "G", and "B" represent red, green, and blue, respectively. For the 1st row, "Spherically Asymmetric Separation (3 Cell Types)", the "Cell Type 3" is defined as a cell that originally belongs to "Cell Type 1" but finally contacts at least one cell in "Cell Type 2", so the mechanical properties of "Cell Type 3" is same to the ones of "Cell Type 1".

The first condition is to program two types of cells, which have strong adhesion within either cell type and weak adhesion between the two cell types (Table 2). Since the cells of either cell type tend to be near each other, the system would separate into two major regions corresponding to the two cell types (2nd row in Fig. 6; Movie S3). If a cell in Type 1 is additionally programmed to differentiate into another fate (Type 3) when contacting a cell of Type 2, a nested structure formed by three cell types appears as seen *in vivo* (1st row in Fig. 6; Movie S4). The second condition is to program two types of cells as well, but the cells in Type 1 and Type 2 have strong and weak adhesion respectively (Table 2), which generates a layered structure in line with the experimental one (3rd row in Fig. 6; Movie S5). Furthermore, we adopt the final state of the layered structure and test if the self-repairing (by removal of half of the cells) and dissociation (by eliminating cell-cell adhesion) take place when the experimental conditions are mimicked *in silico*. As expected, both morphogenetic phenomena are successfully reproduced in simulation (4th and 5th rows in Fig. 6; Movies S6-S7).

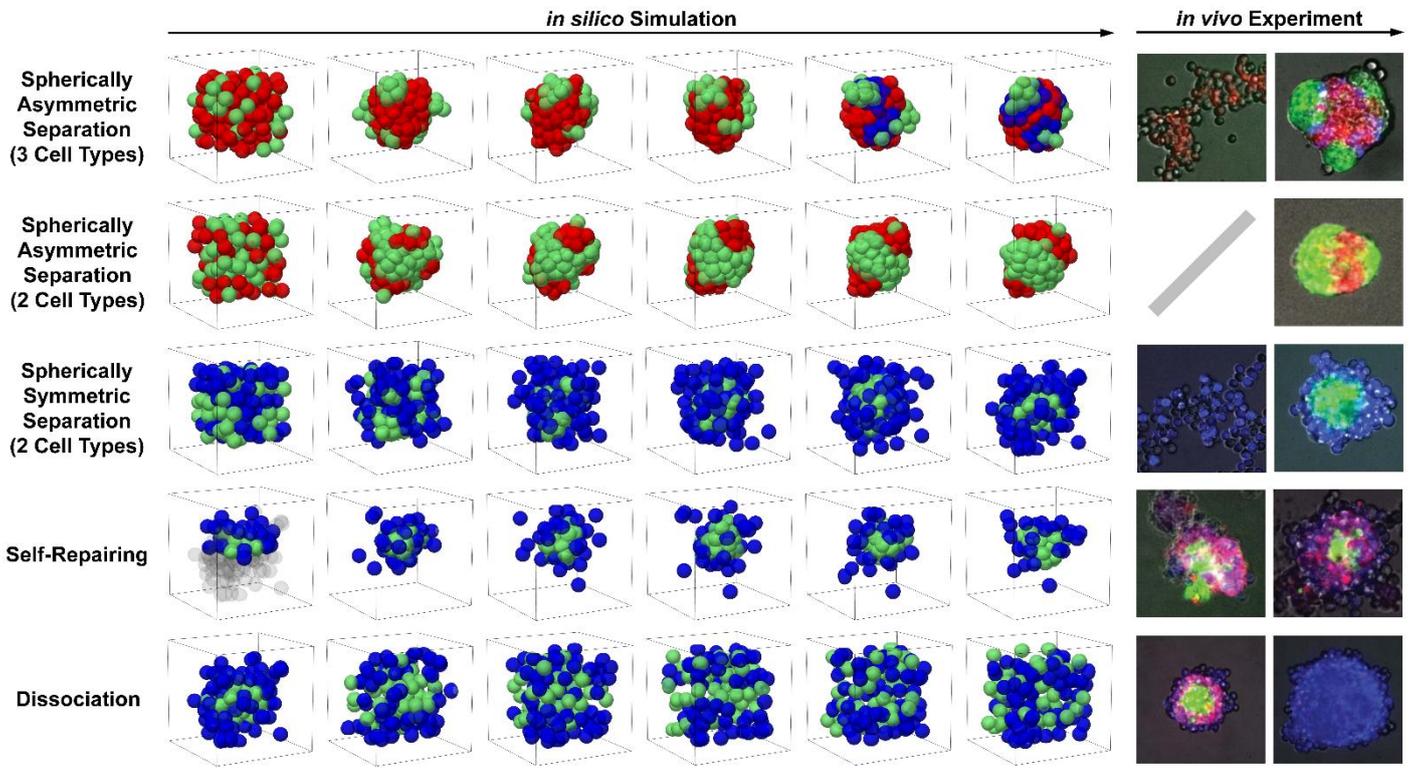

**Fig. 6.** *In silico* simulation (left panel) and *in vivo* experiment (right panel) of the *synNotch* systems. In each row, the *in silico* structures are shown from time points 0 to 50000 in a step of 10000; the computational domain is outlined with a black cube. In the 1st row, since *in silico* time = 40000 (the 5th column), a part of red cells, which contact at least one green cell, are painted blue; in both the 4th and 5th rows, the first *in silico* structure is adopted from the last one in the 3rd row; for the former one, the cells with $z < 0$ or without contact to the cell aggregate are removed and shown by gray shadow. For the *in silico* structures, the cell types and corresponding colors and value assignments on adhesion are listed in Table 2. For the *in vivo* structures, the fluorescent colors are in line with the ones used for the *in silico* structures, except that the red cells in the 4th and 5th rows also correspond to the blue ones *in silico* as they are the same in cell adhesion program. The images of *in vivo* experiment are from [17], reprinted with permission from AAAS.

*3.3. Simulating and exploring more biological processes extensively with MorphoSim*

As exemplified by simulations on nematode embryogenesis and *synNotch* systems, *MorphoSim* is capable of modeling organic and embryonic morphologies precisely. Despite that only the cell-cell attraction is studied as a variable, all physical parameters can be changed and adapted to specific biological processes. The key physical parameters and their biological significance are listed in Table 3, in which we explain how they should be customized for different biological scenes.

Table 3. Tunable physical parameters with specific biological significance.

| Physical Parameter | Biological Significance | Remark |
|---|---|---|
| $\gamma$ | Cell Surface Tension | The cell surface tension is mainly attributed to cortical contractility [43]. As cell surface tension minimizes the cell surface area and pushes a cell to be spherical instead of being deformed, it also characterizes the effect of cell stiffness (hard or soft) if the cell's natural shape is spherical (Fig. S4(a)-(c) and Table S4). In reality, the cell stiffness is contributed by the cytoskeleton and nucleus and varies from cell type to cell type. For instance, the tumor cells are softer than their coexisting benign counterparts [44-46]. |
| $\phi_e$ | Eggshell/Boundary | The size and shape of the eggshell or boundary can be customized according to the system of interest (e.g., ellipsoidal or cylindroid for *C. elegans* embryo and boundless for *synNotch* system) [17,47,48]. |
| $\sigma$ | Cell-Cell Attraction | The attraction between cells usually comes from cell-cell adhesion (e.g., cadherin family), gap junction (e.g., connexin family), and ligand-receptor binding [49-51]. Asymmetric intercellular attraction ($\sigma_{i,j} \neq \sigma_{j,i}$) could be employed to model the biological phenomena in which one cell is unidirectionally attracted by another, such as the chemotactic cell movement induced by cell-cell signaling and engulfment during apoptosis [52-54]. |
| $V_i(t)$ | Cell Volume | The cell volume can be time-dependent according to specific situations. For example, it decreases during cell apoptosis and increases during cell growth [39,55]. The cell volume can also be affected by extrinsic factors like osmotic pressure and substrate stiffness [56,57]. |
| $\tau$ | Ambient Viscosity | The viscosity is dependent on the real scenario [58]. When the environment disobeys the overdamping condition, one can rewrite Eq. (5) accordingly [59,60] |
| $\kappa$ | Stochastic Noise | The stochasticity in cell movement can be raised by intrinsic (e.g., thermal effect and fluctuation) and extrinsic (e.g., the stirring during incubation of *synNotch* spheroid) reasons [17,61,62]. |

## 4. Conclusion and discussion

Multicellular morphology is a fascinating topic and a long-term focus of biological research. Extending on our previous phase field model established with high-quality *in vivo* data, in this work we developed an efficient and scalable *MorphoSim* framework for multicellular systems. We proposed a stabilized numerical scheme and new volume constriction to lower the model's computational cost for simulating a large number of cells simultaneously with considerable accuracy. The *MorphoSim* provides an efficient and powerful tool that not only affords large-scale simulations, but also allows high-dimensional parameter scanning on GPU/CPU clusters. Moreover, the morphodynamics of tricomponent and biocomponent *synNotch* systems can be reproduced by *MorphoSim* in ~ half a day, demonstrating the framework's validity, efficiency, and applicability.

In synthetic biology, more and more attentions are being paid to cell-based morphology synthesis including organoid, embryoid, bio-robot, etc. Taking the *synNotch* system as an example, they are genetically programmed with differential cell adhesion to achieve different modes of structure and function. However, it still faces some challenges: e.g., 1. How do the biological parameters like cell stiffness and surface tension affect the dynamics? (Fig. S4 (a)-(c) and Table S4); and 2. How to choose the parameter combination to optimize a specific function? It would be very helpful if the parameter space can be efficiently explored [63]. *MorphoSim* can regenerate the *synNotch* dynamics quickly and large-scale simulations can also be carried out on GPU and CPU clusters for both mechanistic studies and parameter optimization. Note that the *MorphoSim* parameters (e.g., adhesive strength, motional noise, cell number, spatial resolution, temporal resolution) should be readjusted to the real experimental condition for better simulation performance.

The proposed phase-field framework is also practical to model natural systems such as a developing embryo with large cell numbers. By comparing the *in silico* and *in vivo* morphologies, the low-cost parameter scanning permits the inference of a multicellular system's mechanical state, which is hard to measure directly or infer by the previous phase field model [1]. As the phase field model describes a cell on a dense mesh, more space-related biochemical and biophysical details (e.g., polarity and cytoskeleton) can be added to reconstruct the real system comprehensively.

Despite that *MorphoSim* can simulate over one hundred interacting cells, there is still a great need from the field to keep increasing the cell number and decreasing the computational cost. For example, the early *Drosophila* embryogenesis involves thousands of cells and was simulated using the coarse-grained model or vertex model [64,65]. On the one hand, the limited GPU memory hampers the mesh enlargement/densification when considering a large number of cells. Algorithms like parallel computation [66], adaptive mesh refinement [67], and moving mesh method [68] can be employed to reduce the computational cost and improve the efficiency.

In addition to the algorithmic improvements above, the *MorphoSim* framework can be further developed based on the multiscale method. Instead of using a phase field variable to represent a single cell, we may use a macroscopic phase field variable to represent a group of cells that have the same biophysical properties and microscopic phase field variables to mimic the cells with distinct properties, which could greatly reduce the computational cost and allow for simulations with a much larger number of cells. Besides, for an *in vivo* system with known morphological information (e.g., *C. elegans* embryo) [42], it is feasible to only simulate a special group of cells or the regions of interest instead of the entire system while assigning the cell morphology measured experimentally onto the other cells as a boundary constraint.

### Acknowledgments

We thank Prof. Lei-Han Tang, Prof. Zhongying Zhao, Prof. Xiaojing Yang, and Dr. Zhen Xu for helpful discussions and comments. This work was supported by the National Key R&D Program of China (Grant No. 2021YFF1200500), the National Natural Science Foundation of China (Grant Nos. 12050002, 12090053, 32088101).

### Author contributions

X.K. and G.G. designed and implemented the *MorphoSim* framework, performed simulation and data analysis, and wrote the paper. L.Z. and C.T. conceived and coordinated the study, provided theoretical guidance, and revised the paper.

### Competing interests

The authors declare no competing financial or non-financial interests.

### Data availability

All relevant data are within the paper and its Supplementary Material files. The scripts of *MorphoSim* and its Matlab-based GUI are deposited in https://github.com/XiangyuKuang/MorphoSim.git.

### Code availability

In addition to *MorphoSim*, all other source codes used in this paper are available on reasonable request.

**Supplementary Material**

*MorphoSim: An efficient and scalable phase-field framework for accurately simulating multicellular morphologies*

*Xiangyu Kuang[1†], Guoye Guan[1†], Chao Tang[1,2,3*], Lei Zhang[1,4*]*

[1] Center for Quantitative Biology, Peking University, Beijing 100871, China

[2] Peking-Tsinghua Center for Life Sciences, Peking University, Beijing 100871, China

[3] School of Physics, Peking University, Beijing 100871, China

[4] Beijing International Center for Mathematical Research, Peking University, Beijing 100871, China

[†] These authors contributed equally to this work.

[*] For correspondence: tangc@pku.edu.cn (CT), zhangl@math.pku.edu.cn (LZ)

*Supplemental Figures*

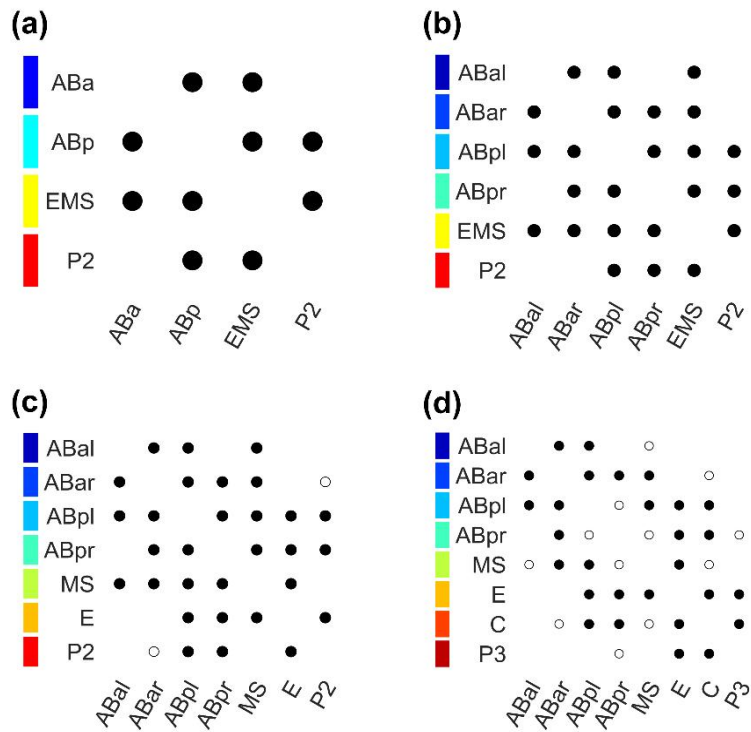

**Fig. S1.** Cell-cell contact map of *C. elegans* embryo at (a) 4-cell, (b) 6-cell, (c) 7-cell, and (d) 8-cell stages. Solid and empty circles denote the conserved and unconserved cell-cell contact pairs, while the remaining blank regions mean that no contact is found between the corresponding cell pair. The contact information is summarized using 17 embryo samples at their last imaging time points of each stage, where the cells undergoing cytokinesis (i.e., with two nuclei embedded by one membrane) are all excluded [Cao et al, *Nat. Commun.*, 2020].

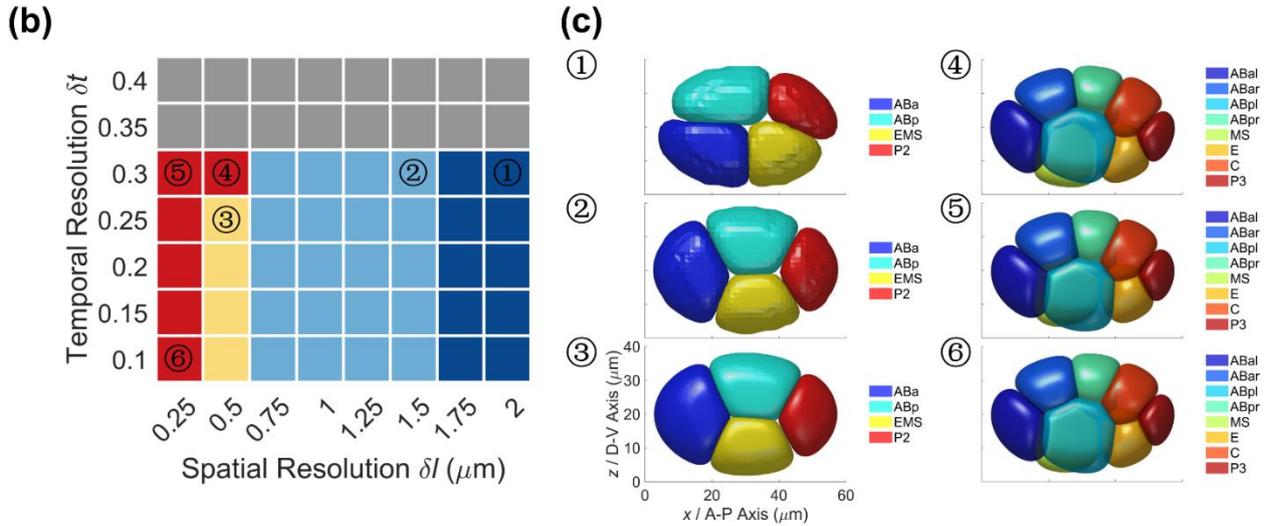

**Fig. S2.** Framework improvement by refinement of spatial and temporal resolutions. (a) Quality-control pipeline to verify if a set of simulations capture the *in vivo* system with enough accuracy. The quantitative criteria for each requirement are introduced in Supplementary Note 1. (b) Simulation results under different spatial and temporal resolutions. Each color represents the stage when the simulation dissatisfies the requirements as shown in (a), except that the red means the simulation passes all the requirements. (c) Final embryonic morphologies under the combinations of spatial and temporal resolutions ① ~ ⑥ labeled in (b).

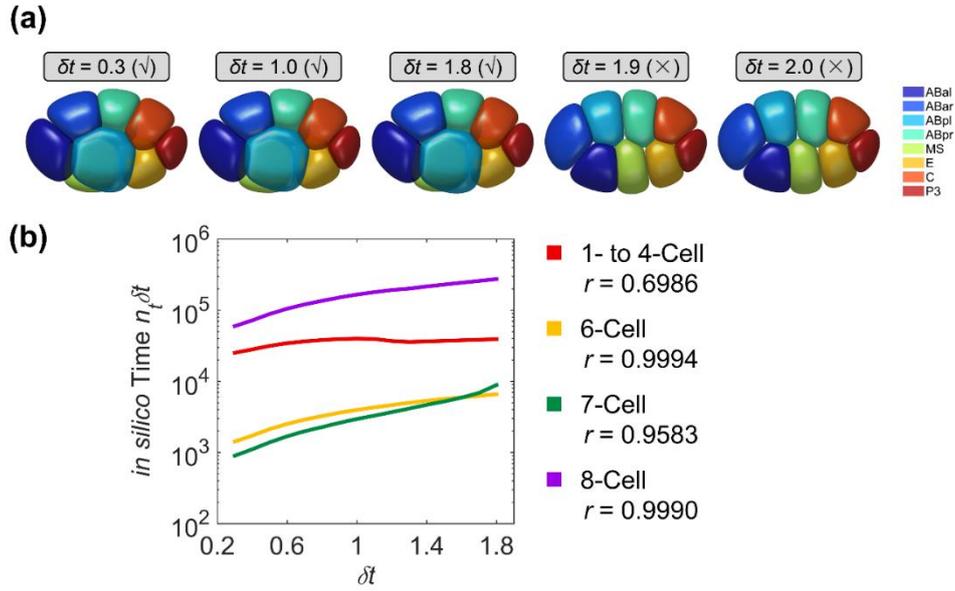

**Fig. S3.** Limitation of the framework with first-order stabilization term. (a) Embryonic morphologies at the end of 8-cell stage under $\delta t$ = 0.3, 1.0, 1.8, 1.9, and 2.0. The ones that successfully achieve the typical structure *in vivo* are labeled with "√" while the others are labeled with "×". (b) *In silico* time $n_t \delta t$ at different stages against temporal resolution $\delta t$. The cell stages and corresponding colors and correlation coefficients $r$ between $\delta t$ and logarithmic $n_t \delta t$ are listed on right.

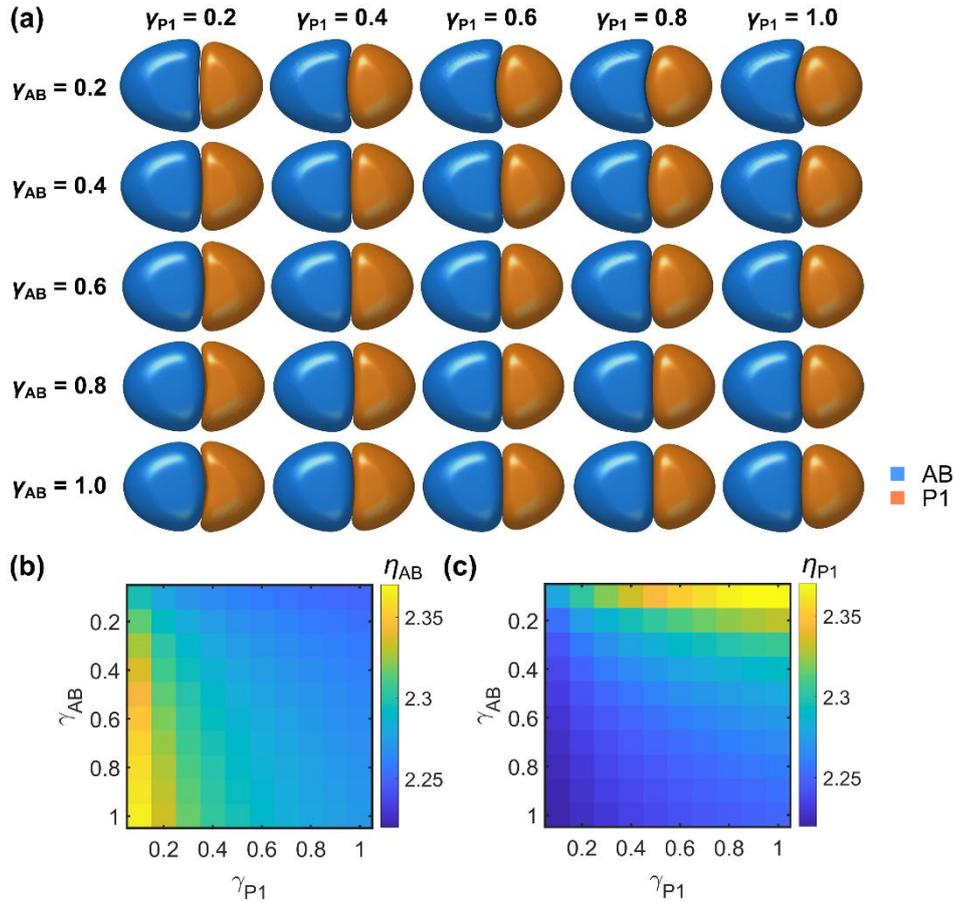

**Fig. S4.** The morphological effect of $\gamma$ that characterizes both cell surface tension and cell stiffness, illustrated by the 2-cell stage with different value assignments. (a) Morphology of AB (blue) and P1 (orange) cells. (b, c) The irregularity score $\eta = \dfrac{Cell\ Surface\ Area^{\frac{1}{2}}}{Cell\ Volume^{\frac{1}{3}}}$ of AB and P1 cells, respectively [Cao et al, *Nat. Commun.*, 2020].

*Legends of Supplemental Movies*

**Movie S1.** Phase-field simulation of *C. elegans* embryogenesis from 12- to 24-cell stages, before modifying the formulation of cell volume constriction. The cell disappearance occurs in the P4 cell at 24-cell stage, as shown in Fig. 3(a).

**Movie S2.** Phase-field simulation of *C. elegans* embryogenesis from 1- to 102-cell stages, after modifying the formulation for cell volume constriction. No cell disappearance occurs throughout the simulation, as shown in Fig. 5(b).

**Movie S3.** Phase-field simulation of the *synNotch* system with spherically asymmetric separation (2 cell types), corresponding to the 2$^{nd}$ row in Fig. 6. The cell types and corresponding colors and value assignments on adhesion are listed in Table 2. The simulation lasts for an *in silico* time = 0 ~ 50000.

**Movie S4.** Phase-field simulation of the *synNotch* system with spherically asymmetric separation (3 cell types), corresponding to the 1$^{st}$ row in Fig. 6. The cell types and corresponding colors and value assignments on adhesion are listed in Table 2. The simulation lasts for an *in silico* time = 0 ~ 50000; since *in silico* time = 40000, the red cells which contact at least one green cell are painted blue.

**Movie S5.** Phase-field simulation of the *synNotch* system with spherically asymmetric separation (2 cell types), corresponding to the 3$^{rd}$ row in Fig. 6. The cell types and corresponding colors and value assignments on adhesion are listed in Table 2. The simulation lasts for an *in silico* time = 0 ~ 50000.

**Movie S6.** Phase-field simulation of the *synNotch* system with self-repairing, corresponding to the 4$^{th}$ row in Fig. 6. The cell types and corresponding colors and value assignments on adhesion are listed in Table 2. The initial state is adopted from the final state in the simulation of spherically asymmetric separation (2 cell types) (3$^{rd}$ row in Fig. 6; Movie S5), where the cells with $z < 0$ or without contact to the cell aggregate are removed. The simulation lasts for an *in silico* time = 0 ~ 50000.

**Movie S7.** Phase-field simulation of the *synNotch* system with dissociation (2 cell types), corresponding to the 5$^{th}$ row in Fig. 6. The cell types and corresponding colors and value assignments on adhesion are listed in Table 2. The initial state is adopted from the final state in the simulation of spherically asymmetric separation (2 cell types) (3$^{rd}$ row in Fig. 6; Movie S5). The simulation lasts for an *in silico* time = 0 ~ 50000.

*Supplementary Note 1*

**Search of the minimal spatial and temporal resolutions enough for reproducing the previous findings**

Here, we attempt to minimize the spatial and temporal resolutions while guaranteeing the results that are consistent with the ones in [Kuang et al, *PLoS Comput. Biol.*, 2022]. The step-by-step simulation procedure and key biological findings of the original framework are summarized in Fig. S2(a). The requirements that judge if the framework works precisely enough from the 1- to 8-cell stages mainly include 3 parts: 1. the cell-cell contact maps are the same as the ones conserved between individual embryos; 2. the previously reported cell-cell adhesion programs can be inferred by comparison to experiment and parameter scanning; 3. the embryonic morphologies resemble the ones *in vivo*. The quantitative criteria and search pipeline are introduced detailly below.

*1.1. Grid setting*

We predetermine the computational domain $\Omega$ as a cuboid with a size of $(L_x, L_y, L_z) = (60 \text{ μm}, 40 \text{ μm}, 30 \text{ μm})$. Then, we consider the scanning range of space step (spatial resolution) as $\delta l = 0.25: 0.25: 2.00$ (μm) and time step (temporal resolution) as $\delta t = 0.10: 0.05: 0.40$. For each value of $\delta l$, the grid points of a mesh in three orthogonal directions are $n_x = \text{CEIL}\left(\frac{L_x}{\delta l}\right), n_y = \text{CEIL}\left(\frac{L_y}{\delta l}\right), n_z = \text{CEIL}\left(\frac{L_z}{\delta l}\right)$, where CEIL is the round-up integral function.

*1.2. Definition of stable state*

For a phase field $\phi_i(r, t)$ of a specific cell $i$ and at a specific time point $t$, its centroid is calculated as $r_i^t = \frac{\int_\Omega r \phi_i^t \mathrm{d}r}{\int_\Omega \phi_i^t \mathrm{d}r}$. We use the consecutive $\text{FLOOR}\left(\frac{10}{\delta t}\right)$ time steps to calculate its average velocity during this interval, namely, $v_i^t = \frac{r_i^t - r_i^{t - \text{FLOOR}\left(\frac{10}{\delta t}\right)\delta t}}{10}$, where FLOOR is the round-down integral function. When the root-mean-square velocity of all $N$ cells inside the eggshell, $\overline{v^t} = \sqrt{\frac{1}{N}\sum_{i=1}^{N} |v_i^t|^2}$, is smaller than $1 \times 10^{-4}$, the whole system is regarded as reaching a stable state.

*1.3. Regeneration of the key findings in the previous framework*

The step-by-step establishment of the original phase-field framework revealed a series of significant biological processes, which are in favorable agreement with *in vivo* observations [Kuang et al, *PLoS Comput. Biol.*, 2022]. The degradation of spatial and temporal resolutions is requested to remain the previously reported key findings unchanged, otherwise, the resolution is considered not precise enough. The simulation procedures (labeled by capital letters) and quantitative criteria (labeled by solid circles) are listed below.

A. Simulating the 1- to 2-cell stages without cell-cell attraction, until reaching the stable state respectively.
   - No numeric overflow occurs.
B. Simulating the 3-cell stage without cell-cell attraction, until reaching the stable state.
   - The ABa-P1 contact is established.
C. Simulating the 4-cell stage without cell-cell attraction, until reaching the stable state.
   - The ABa-ABp, ABa-EMS, ABp-EMS, ABp-P2, EMS-P2 contacts are established (Fig. S1(a)).
   - The anterior-posterior and dorsal-ventral axes are established.

$$\boldsymbol{n}_{\text{AP}} = \frac{\boldsymbol{r}_{\text{ABa}} - \boldsymbol{r}_{\text{P2}}}{|\boldsymbol{r}_{\text{ABa}} - \boldsymbol{r}_{\text{P2}}|}, \left|\cos^{-1}(\boldsymbol{n}_{\text{AP}} \cdot (1,0,0))\right| < 10°$$

$$\boldsymbol{n}_{\text{DV}} = \frac{\boldsymbol{r}_{\text{ABp}} - \boldsymbol{r}_{\text{EMS}}}{|\boldsymbol{r}_{\text{ABp}} - \boldsymbol{r}_{\text{EMS}}|}, \left|\cos^{-1}(\boldsymbol{n}_{\text{DV}} \cdot (0,1,0))\right| < 10°$$

- The areas of 5 contacts are essentially smaller than the real values. Here, we label the simulated and experimental areas of cell-cell contact $h$ with $S_{\text{sim},h}$ and $S_{\text{exp},h}$ respectively.

$$\text{MEAN}\left\{\frac{S_{\text{sim},h} - S_{\text{exp},h}}{S_{\text{exp},h}}\right\} < 0, \text{ for all 5 contacts}$$

D. Simulating the 4-cell stage with global cell-cell attraction $\sigma = 0: 0.1: 1.5$ and searching for the optimal $\sigma_S$ value with the smallest $\text{MEAN}\left\{\left|\frac{S_{\text{sim},h} - S_{\text{exp},h}}{S_{\text{exp},h}}\right|\right\}$ for ABa-ABp, ABa-EMS, ABp-EMS, ABp-P2 contacts.

- The areas of all 5 contacts are enlarged compared to the ones without cell-cell attraction.

$$\text{MEAN}\left\{\left|\frac{S_{\text{sim},h} - S_{\text{exp},h}}{S_{\text{exp},h}}\right|\right\}_{\sigma \neq 0} < \text{MEAN}\left\{\left|\frac{S_{\text{sim},h} - S_{\text{exp},h}}{S_{\text{exp},h}}\right|\right\}_{\sigma = 0}, \text{ for all 5 contacts}$$

- The areas of ABa-ABp, ABa-EMS, ABp-EMS, ABp-P2 contacts are close to the real values.

$$\left|\frac{S_{\text{sim},h} - S_{\text{exp},h}}{S_{\text{exp},h}}\right| < 0.20, \text{ for ABa} - \text{ABp, ABa} - \text{EMS, ABp} - \text{EMS, ABp} - \text{P2 contacts}$$

- The area of EMS-P2 contact is essentially larger than the real value.

$$\frac{S_{\text{sim},h} - S_{\text{exp},h}}{S_{\text{exp},h}} > 0.40, \text{ for EMS} - \text{P2 contact}$$

E. Simulating the 4-cell stage with lower cell-cell attraction $\sigma = 0: 0.1: (\sigma_S - 0.1)$ specified for EMS-P2 contact and searching for the optimal $\sigma_W$ value with the smallest $\text{MEAN}\left\{\left|\frac{S_{\text{sim},h} - S_{\text{exp},h}}{S_{\text{exp},h}}\right|\right\}$ for all 5 contacts.

- The areas of all 5 contacts are close to the real values.

$$\left|\frac{S_{\text{sim},h} - S_{\text{exp},h}}{S_{\text{exp},h}}\right| < 0.20, \text{ for all 5 contacts}$$

F. Simulating the 6-cell stage to its quasi-steady state $\left(\frac{d\bar{v}}{dt}\bigg|_{t=t_q} = 0, \frac{d^2\bar{v}}{dt^2}\bigg|_{t=t_q} > 0\right)$

- The cell-cell contact map is the same as the one generated by the original framework and covers all conserved contacts and non-contacts observed experimentally (Fig. S1(b)).

G. Simulating the 7-cell stage to its quasi-steady state $\left(\frac{d\bar{v}}{dt}\bigg|_{t=t_q} = 0, \frac{d^2\bar{v}}{dt^2}\bigg|_{t=t_q} > 0\right)$

- The cell-cell contact map is the same as the one generated by the original framework and covers all conserved contacts and non-contacts observed experimentally (Fig. S1(c)).

H. Simulating the 8-cell stage for a long enough duration (*in silico* time = step number × step length = 10000)

- The simulation with weak attraction in ABpl-E contact $(\sigma_{\text{ABpl, E}} = \sigma_W)$ generates the typical 8-cell structure stabilized in 3-dimensional, whose cell-cell contact map is the same as the one generated by the original framework and covers all conserved contacts and non-contacts observed experimentally (Fig. S1(d)).
- The simulation with strong attraction in ABpl-E contact $(\sigma_{\text{ABpl,E}} = \sigma_S)$ is unstabilized and collapses into an incorrect structure much earlier than the one with weak attraction.

*1.4. Search result*

Given the original spatial resolution $\delta l = 0.25$ (μm) and temporal resolution $\delta t = 0.10$, we repeat the simulation procedure from 1- to 8-cell stages under $\delta l = 0.25$: $0.25$: $2.00$ (μm) and $\delta t = 0.10$: $0.05$: $0.40$ as shown in Fig. S2(a). The experimental data of cell division order and axis and volume segregation ratio in *C. elegans* early embryogenesis is obtained from [Cao et al, *Nat. Commun.*, 2020] and inputted. The eggshell boundary is set as a fitted ellipsoid with semi-axes $L_x = 28.28$ μm, $L_y = 12.87$ μm, $L_z = 18.81$ μm [Moshtagh, *MATLAB Central File Exchange, 2022*], while the region $|y| > w_{max} = 10.21$ μm is removed to resemble the lateral compression in fluorescent imaging; the $x$, $y$, and $z$ coordinates of the eggshell center are prescribed as $(30\ \mu m, 15\ \mu m, 20\ \mu m)$, and the eggshell structure is shown in Fig. 1(a). The stage when a simulation fails to meet the requirement is shown in Fig. S2(b) and (c), revealing the largest degraded resolutions $\delta l = 0.50$ (μm) and $\delta t = 0.30$ that are still at work (Combination ④ in Fig. S2(b) and (c)). The fitted attraction intensities in the EMS-P2 interface and the others are fitted as $\sigma_W = 0$ and $\sigma_S = 0.5$ respectively.

*Supplementary Note 2*

**Iterative process of *MorphoSim***

Given the initial conditions $\phi_i^0$, $\phi_i^{n+1}$ with $n = 0$ (i.e., $\phi_i^1$) is computed by the first-order semi-implicit scheme:

$$\frac{\tau}{\delta t}(\phi_i^{n+1} - \phi_i^n) = \gamma \Delta \phi_i^{n+1} + F_i^n - S(\phi_i^{n+1} - \phi_i^n), \tag{S1}$$

where $F_i^n$ is the nonlinear term and expressed by:

$$F_i^n = -\gamma c\left(4(\phi_i^n)^3 - 6(\phi_i^n)^2 + 2\phi_i^n\right) - g_e \phi_i^n \phi_e^2 - g\phi_i^n \sum_{j \neq i}^N (\phi_j^n)^2 - \nabla\phi_i^n \cdot \sum_{j \neq i}^N \sigma_{i,j} \nabla\phi_j^n + M'\left(1 - \frac{\int_\Omega \phi_i^n \mathrm{d}\boldsymbol{r}}{V_i(t)}\right)|\nabla\phi_i^n|, \tag{S2}$$

In the simulations of *synNotch* systems, Gaussian white noise is added to the evolution equation. Thus, the nonlinear term $F_i^n$ turns into:

$$\begin{aligned}F_i^n =& -\gamma c\left(4(\phi_i^n)^3 - 6(\phi_i^n)^2 + 2\phi_i^n\right) - g_e \phi_i^n \phi_e^2 - g\phi_i^n \sum_{j \neq i}^N (\phi_j^n)^2 - \nabla\phi_i^n \cdot \sum_{j \neq i}^N \sigma_{i,j} \nabla\phi_j^n + M'\left(1 - \frac{\int_\Omega \phi_i^n \mathrm{d}\boldsymbol{r}}{V_i(t)}\right)|\nabla\phi_i^n| \\ & - \kappa\sqrt{\delta t}\nabla\phi_i^n \cdot u_i^n,\end{aligned} \tag{S3}$$

where $u_i^n$ is the velocity of cell $i$'s random motion sampled from the standard Gaussian distribution at each time step. After $\phi_i^0$ and $\phi_i^1$ are obtained, $\phi_i^{n+1}$ with $n \geq 1$ is computed by the second-order semi-implicit scheme:

$$\frac{\tau}{2\delta t}\left(3\phi_i^{n+1} - 4\phi_i^n + \phi_i^{n-1}\right) = \gamma \Delta \phi_i^{n+1} + 2F_i^n - F_i^{n-1} - S(\phi_i^{n+1} - 2\phi_i^n + \phi_i^{n-1}). \tag{S4}$$

*Supplementary Note 3*

**User guidebook for *MorphoSim* GUI on Matlab**

*2.1. Brief introduction*

    The graphical user interface (GUI) of *MorphoSim* (Morphology Simulator) is constructed using Matlab 2018b. One can utilize this GUI to compute the morphological change of a multicellular system conveniently.

*2.2. Use flow*

1) Download the Folder "MorphoSim" from GitHub https://github.com/XiangyuKuang/MorphoSim.git (Fig. G1).
2) Open the MorphoSim.m in Matlab and click "Run", then an interactive interface appears (Fig. G2).
   - The Matlab version should be no earlier than 2018b.
   - Click "Set Path" to add the Folder "MorphoSim" with its subfolders before running the programs (Fig. G3).
3. To compute a multicellular system with specific cell names and attraction matrix, use the upper panel:
A. Import the information of cell name and attraction matrix (Fig. G4).
   - The file format is shown in "Example\Example_CellName&AttractionMatrix_*.xls", where "*" is "1" or "2".
   - The number in Row *j* and Column *i* means the attraction ($\sigma_{i,j}$) from Cell *j* to Cell *i*.
   - The file "Example\Example_CellName&AttractionMatrix_1.xls" contains the matrix with strong attraction in ABpl-E contact (Fig. G4(b) and (e)). The file "Example\Example_CellName&AttractionMatrix_2.xls" contains the matrix with weak attraction in ABpl-E contact (Fig. G4(c) and (f)).
B. Import the initial cell pattern which is kept as binary regions and can be designated arbitrarily.
   - The file format is shown in "Example\Example_8Cell.mat".
   - The cellular regions should be set inside the eggshell shown in "Example\Example_Eggshell.mat", in which the interior and exterior of the eggshell are labeled by 0 and 1 respectively.
C. Set the *in silico* time, step length ($\delta t$; recommended value: 2), and saving interval for computation (Fig. G4(a)).
D. Choose if the eggshell is considered.
E. Give a name like "Test" for the simulation and its output storage.
F. Click "Run", then a folder named by Step E is generated and the cell patterns at spaced time points would be saved as "Test\Test_*.mat", where "*" denotes the *in silico* time corresponding to each file (Fig. G5).
   - One can know the progress through the files saved intermittently.
4. To illustrate a multicellular structure outputted by *MorphoSim*, use the lower panel (Fig. G4(b) and (c)):
A. Import the information of cell names.
   - The file format is shown in "Example\Example_CellName.xls".
B. Import the selected cell pattern outputted by *MorphoSim*.
C. Click "Plot", then a figure illustrating the multicellular structure would be shown (Fig. G4(d)-(f)).

*2.3. Contact*

    All the fundamental scripts of the *MorphoSim* GUI have been uploaded onto Github https://github.com/XiangyuKuang/MorphoSim.git. If there is any question, please contact Xiangyu Kuang (kuangxy@pku.edu.cn) or Guoye Guan (guanguoye@gmail.com).

*2.4. Figures for the user guidebook*

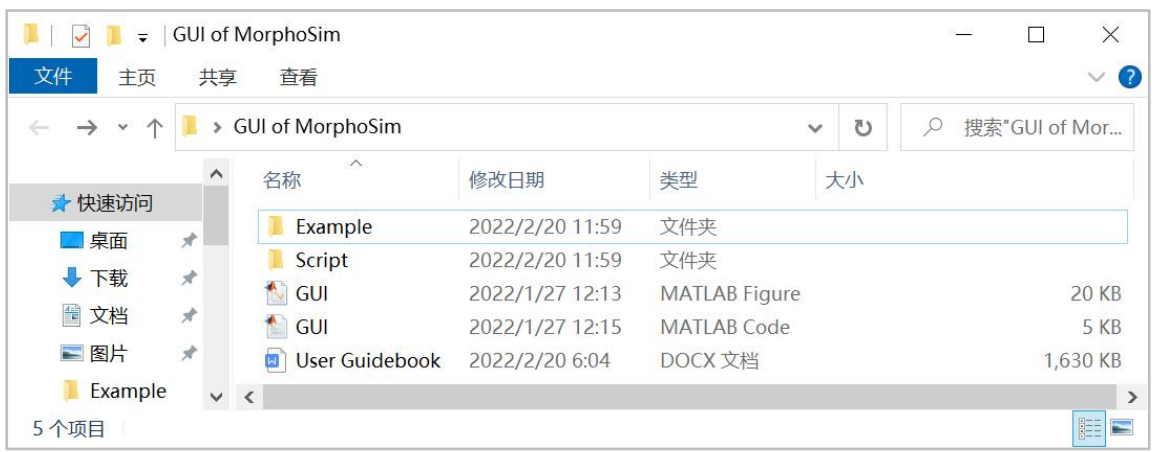

**Fig. G1.** The files and subfolders in the Folder "MorphoSim" downloaded from GitHub.

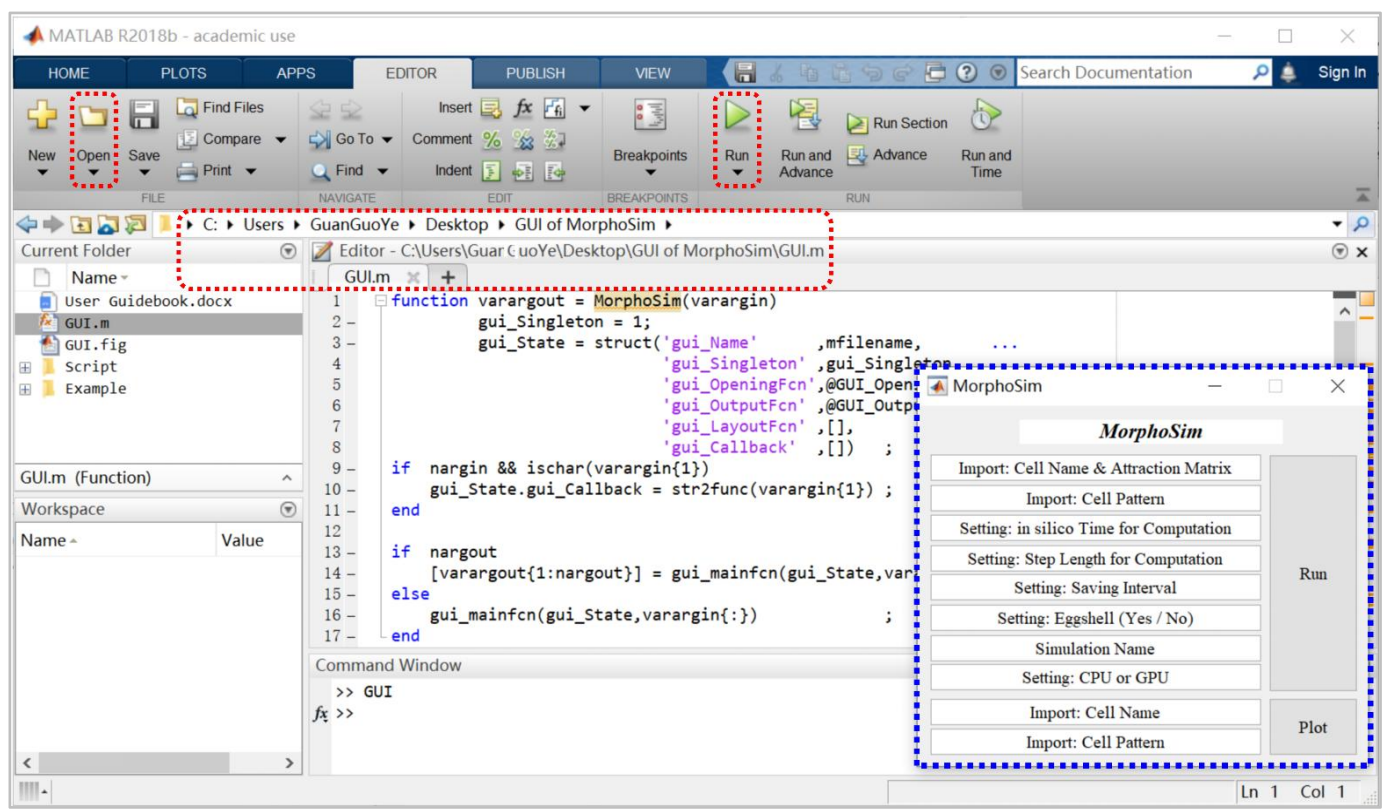

**Fig. G2.** Instruction to open the *MorphoSim* GUI. The script imported and the buttons for loading and running it are noted by dashed red frames; the interface of *MorphoSim* generated after clicking "Run" is noted by a dashed blue frame.

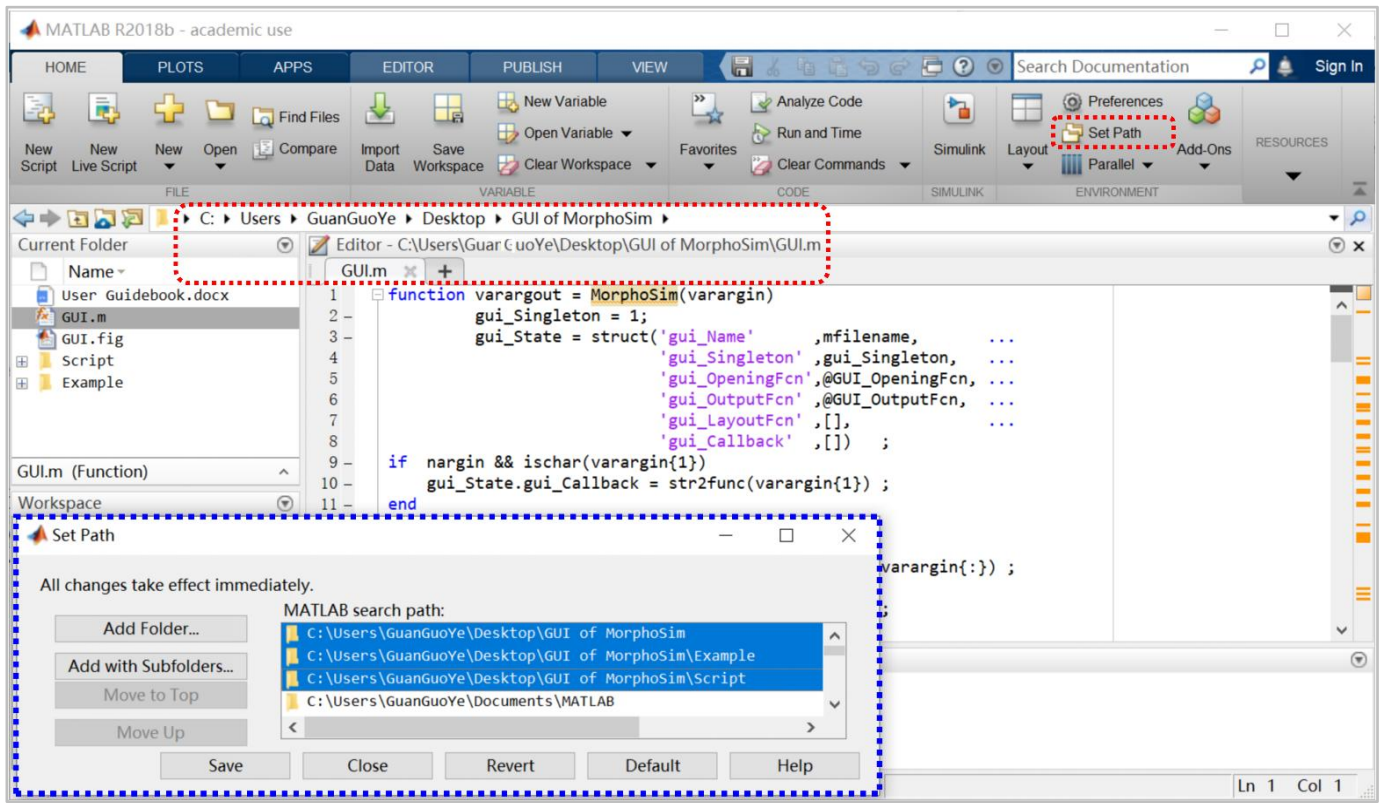

**Fig. G3.** Instruction to set the path for the *MorphoSim* GUI. The script imported and the button for setting path are noted by dashed red frames; the interface for setting path generated after clicking "Set Path" is noted by a dashed blue frame.

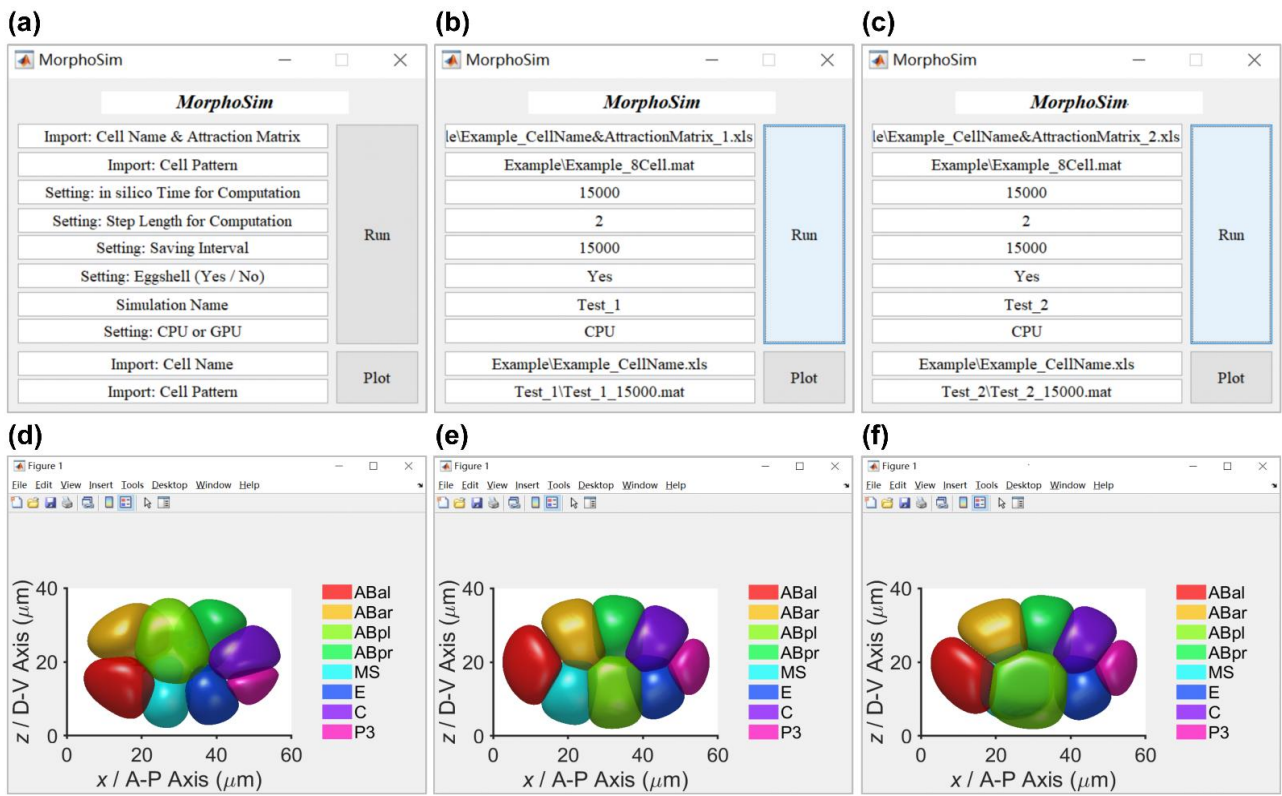

**Fig. G4.** The graphical user interface of *MorphoSim*. (a) The interface and instruction of inputs required. (b)(c) The simulation inputs for 8-cell *C. elegans* embryogenesis with strong and weak adhesion in ABpl-E contact respectively. (d) The initial state (*in silico* time = 0) of the 8-cell embryo. (e)(f) The final states (*in silico* time = 15000) of the 8-cell embryo with strong and weak adhesion in ABpl-E contact respectively. Note that this whole figure is the same as Fig. 4 in the main text.

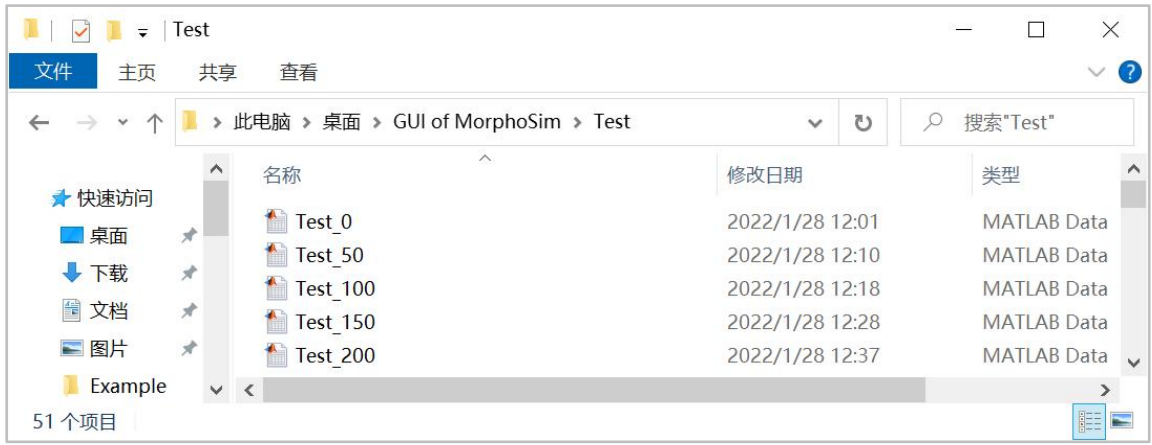

**Fig. G5.** The files outputted and saved intermittently.